\documentclass[%
reprint,
superscriptaddress,
amsmath,amssymb,
aps,
prd,
noeprint,
floatfix,
]{revtex4-2}
\usepackage{ragged2e}
\usepackage[dvipsnames]{xcolor}
\usepackage{graphicx}% Include figure files
\usepackage{dcolumn}% Align table columns on decimal point
\usepackage{bm}% bold math
\long\def\/*#1*/{}
\usepackage[normalem]{ulem}
\usepackage{xcolor}          
\usepackage{indentfirst} 

\makeatletter
\let\orig@fnsymbol\@fnsymbol
\def\@fnsymbol#1{\ifcase#1\or a\or b\or c\or d\or e\or f\or g\or h\or i\or j\or
                 k\or l\or m\or n\or o\or p\or q\or r\or s\or t\or u\or v\or w\or x\or y\or z\else\@ctrerr\fi}
\makeatother

\begin{document}

\title{First Measurement of Neutrino Emissions from Spent Nuclear Fuel by the Double Chooz Experiment}
% === Double Chooz — PRL author/affiliation block (standard; present address via \thanks) ===

\author{T.~Abrah\~{a}o}
\affiliation{Centro Brasileiro de Pesquisas F\'{i}sicas, Rio de Janeiro, RJ, 22290-180, Brazil}
\affiliation{APC, Universit{\'e} de Paris, CNRS, Astroparticule et Cosmologie, F-75006, Paris}

\author{H.~Almazan}
\thanks{Now at: Donostia International Physics Center (DIPC), BERC Basque Excellence Research Centre, Donostia, Spain}
\affiliation{Max-Planck-Institut f\"{u}r Kernphysik, 69117 Heidelberg, Germany}

\author{J.C.~dos Anjos}
\affiliation{Centro Brasileiro de Pesquisas F\'{i}sicas, Rio de Janeiro, RJ, 22290-180, Brazil}

\author{S.~Appel}
\affiliation{Physik Department, Technische Universit\"{a}t M\"{u}nchen, 85748 Garching, Germany}

\author{J.C.~Barriere}
\affiliation{IRFU, CEA, Universit\'{e} Paris-Saclay, 91191 Gif-sur-Yvette, France}

\author{I.~Bekman}
\affiliation{III. Physikalisches Institut, RWTH Aachen University, 52056 Aachen, Germany}

\author{T.J.C.~Bezerra}
\thanks{Now at: Department of Physics and Astronomy, University of Sussex, Falmer, Brighton, UK}
\affiliation{Subatech, CNRS, Universit\'{e} de Nantes, IMT-Atlantique, 44307 Nantes, France}

\author{L.~Bezrukov}
\noaffiliation

\author{E.~Blucher}
\affiliation{The Enrico Fermi Institute, The University of Chicago, Chicago, Illinois 60637, USA}

\author{C.~Bourgeois}
\affiliation{IJC Laboratory, CNRS, Universit\'e Paris-Saclay, Orsay, France}

\author{C.~Buck}
\affiliation{Max-Planck-Institut f\"{u}r Kernphysik, 69117 Heidelberg, Germany}

\author{J.~Busenitz}
\affiliation{Department of Physics and Astronomy, University of Alabama, Tuscaloosa, Alabama 35487, USA}

\author{A.~Cabrera}
\affiliation{APC, Universit{\'e} de Paris, CNRS, Astroparticule et Cosmologie, F-75006, Paris}
\affiliation{IJC Laboratory, CNRS, Universit\'e Paris-Saclay, Orsay, France}
\affiliation{LNCA Underground Laboratory, CNRS-CEA, Chooz, France}

\author{M.~Cerrada}
\affiliation{Centro de Investigaciones Energ\'{e}ticas, Medioambientales y Tecnol\'{o}gicas, CIEMAT, 28040, Madrid, Spain}

\author{E.~Chauveau}
\affiliation{Univ. Bordeaux, CNRS, LP2I, UMR 5797, F-33170 Gradignan, France}

\author{P.~Chimenti}
\thanks{Now at: Universidade Estadual de Londrina, 86057-970 Londrina, Brazil}
\affiliation{Centro Brasileiro de Pesquisas F\'{i}sicas, Rio de Janeiro, RJ, 22290-180, Brazil}

\author{O.~Corpace}
\affiliation{IRFU, CEA, Universit\'{e} Paris-Saclay, 91191 Gif-sur-Yvette, France}

\author{J.V.~Dawson}
\affiliation{APC, Universit{\'e} de Paris, CNRS, Astroparticule et Cosmologie, F-75006, Paris}

\author{J.\,F.~Du}
\affiliation{IJC Laboratory, CNRS, Universit\'e Paris-Saclay, Orsay, France}
\affiliation{LNCA Underground Laboratory, CNRS-CEA, Chooz, France}

\author{Z.~Djurcic}
\affiliation{Argonne National Laboratory, Argonne, Illinois 60439, USA}

\author{A.~Etenko}
\noaffiliation

\author{H.~Furuta}
\affiliation{Research Center for Neutrino Science, Tohoku University, Sendai 980-8578, Japan}

\author{I.~Gil-Botella}
\affiliation{Centro de Investigaciones Energ\'{e}ticas, Medioambientales y Tecnol\'{o}gicas, CIEMAT, 28040, Madrid, Spain}

\author{A.~Givaudan}
\affiliation{APC, Universit{\'e} de Paris, CNRS, Astroparticule et Cosmologie, F-75006, Paris}

\author{H.~Gomez}
\affiliation{APC, Universit{\'e} de Paris, CNRS, Astroparticule et Cosmologie, F-75006, Paris}
\affiliation{IRFU, CEA, Universit\'{e} Paris-Saclay, 91191 Gif-sur-Yvette, France}

\author{M.C.~Goodman}
\affiliation{Argonne National Laboratory, Argonne, Illinois 60439, USA}

\author{T.~Hara}
\affiliation{Department of Physics, Kobe University, Kobe, 657-8501, Japan}

\author{J.~Haser}
\affiliation{Max-Planck-Institut f\"{u}r Kernphysik, 69117 Heidelberg, Germany}

\author{D.~Hellwig}
\affiliation{III. Physikalisches Institut, RWTH Aachen University, 52056 Aachen, Germany}

\author{A.~Hourlier}
\affiliation{APC, Universit{\'e} de Paris, CNRS, Astroparticule et Cosmologie, F-75006, Paris}

\author{M.~Ishitsuka}
\thanks{Now at: Tokyo University of Science, Noda, Chiba, Japan}
\affiliation{Department of Physics, Institute of Science Tokyo, Tokyo, 152-8551, Japan}

\author{J.~Jochum}
\affiliation{Kepler Center for Astro and Particle Physics, Universit\"{a}t T\"{u}bingen, 72076 T\"{u}bingen, Germany}

\author{C.~Jollet}
\affiliation{Univ. Bordeaux, CNRS, LP2I, UMR 5797, F-33170 Gradignan, France}

\author{K.~Kale}
\affiliation{Univ. Bordeaux, CNRS, LP2I, UMR 5797, F-33170 Gradignan, France}

\author{M.~Kaneda}
\affiliation{Department of Physics, Institute of Science Tokyo, Tokyo, 152-8551, Japan}

\author{M.~Karakac}
\affiliation{APC, Universit{\'e} de Paris, CNRS, Astroparticule et Cosmologie, F-75006, Paris}

\author{T.~Kawasaki}
\affiliation{Department of Physics, Kitasato University, Sagamihara, 252-0373, Japan}

\author{E.~Kemp}
\affiliation{Universidade Estadual de Campinas-UNICAMP, Campinas, SP, 13083-970, Brazil}

\author{D.~Kryn}
\affiliation{APC, Universit{\'e} de Paris, CNRS, Astroparticule et Cosmologie, F-75006, Paris}

\author{M.~Kuze}
\affiliation{Department of Physics, Institute of Science Tokyo, Tokyo, 152-8551, Japan}

\author{T.~Lachenmaier}
\affiliation{Kepler Center for Astro and Particle Physics, Universit\"{a}t T\"{u}bingen, 72076 T\"{u}bingen, Germany}

\author{C.E.~Lane}
\affiliation{Department of Physics, Drexel University, Philadelphia, Pennsylvania 19104, USA}

% --- T. Lasserre (corresponding) ---
\author{T.~Lasserre\textsuperscript{*}}
%\thanks{*\,Corresponding author: thierry.lasserre@mpi-hd.mpg.de}
\affiliation{APC, Universit{\'e} de Paris, CNRS, Astroparticule et Cosmologie, F-75006, Paris}
\affiliation{IRFU, CEA, Universit\'{e} Paris-Saclay, 91191 Gif-sur-Yvette, France}
\affiliation{Max-Planck-Institut f\"{u}r Kernphysik, 69117 Heidelberg, Germany}

\author{D.~Lhuillier}
\affiliation{IRFU, CEA, Universit\'{e} Paris-Saclay, 91191 Gif-sur-Yvette, France}

\author{H.P.~Lima Jr}
\thanks{Now at: Gran Sasso Science Institute, 67100 L'Aquila, Italy}
\affiliation{Centro Brasileiro de Pesquisas F\'{i}sicas, Rio de Janeiro, RJ, 22290-180, Brazil}

\author{M.~Lindner}
\affiliation{Max-Planck-Institut f\"{u}r Kernphysik, 69117 Heidelberg, Germany}

\author{J.M.~LoSecco}
\affiliation{University of Notre Dame, Notre Dame, Indiana 46556, USA}

\author{B.~Lubsandorzhiev}
\noaffiliation

\author{J.~Maeda}
\affiliation{Department of Physics, Tokyo Metropolitan University, Tokyo, 192-0397, Japan}
\affiliation{Department of Physics, Kobe University, Kobe, 657-8501, Japan}

\author{C.~Mariani}
\affiliation{Center for Neutrino Physics, Virginia Tech, Blacksburg, Virginia 24061, USA}

\author{J.~Maricic}
\thanks{Now at: Physics \& Astronomy Department, University of Hawaii at Manoa, Honolulu, Hawaii, USA}
\affiliation{Department of Physics, Drexel University, Philadelphia, Pennsylvania 19104, USA}

\author{J.~Martino}
\affiliation{Subatech, CNRS, Universit\'{e} de Nantes, IMT-Atlantique, 44307 Nantes, France}

\author{T.~Matsubara}
\thanks{Now at: High Energy Accelerator Research Organization (KEK), Tsukuba, Ibaraki, Japan}
\affiliation{Department of Physics, Tokyo Metropolitan University, Tokyo, 192-0397, Japan}

\author{G.~Mention}
\affiliation{IRFU, CEA, Universit\'{e} Paris-Saclay, 91191 Gif-sur-Yvette, France}

\author{A.~Meregaglia}
\affiliation{Univ. Bordeaux, CNRS, LP2I, UMR 5797, F-33170 Gradignan, France}

\author{T.~Miletic}
\thanks{Now at: Physics Department, Arcadia University, Glenside, PA 19038, USA}
\affiliation{Department of Physics, Drexel University, Philadelphia, Pennsylvania 19104, USA}

\author{R.~Milincic}
\affiliation{Department of Physics, Drexel University, Philadelphia, Pennsylvania 19104, USA}

\author{A.~Minotti}
\affiliation{IRFU, CEA, Universit\'{e} Paris-Saclay, 91191 Gif-sur-Yvette, France}

\author{X.~Mougeot}
\affiliation{Laboratoire National Henri Becquerel, LIST, CEA, 91191 Gif-sur-Yvette Cedex, France}

\author{D.~Navas-Nicol\'as}
\affiliation{Centro de Investigaciones Energ\'{e}ticas, Medioambientales y Tecnol\'{o}gicas, CIEMAT, 28040, Madrid, Spain}
\affiliation{IJC Laboratory, CNRS, Universit\'e Paris-Saclay, Orsay, France}

\author{Y.~Nikitenko}
\affiliation{III. Physikalisches Institut, RWTH Aachen University, 52056 Aachen, Germany}

\author{P.~Novella}
\thanks{Now at: Instituto de F\'{i}sica Corpuscular, IFIC (CSIC/UV), 46980 Paterna, Spain}
\affiliation{Centro de Investigaciones Energ\'{e}ticas, Medioambientales y Tecnol\'{o}gicas, CIEMAT, 28040, Madrid, Spain}

\author{L.~Oberauer}
\affiliation{Physik Department, Technische Universit\"{a}t M\"{u}nchen, 85748 Garching, Germany}

\author{M.~Obolensky}
\affiliation{APC, Universit{\'e} de Paris, CNRS, Astroparticule et Cosmologie, F-75006, Paris}

% --- A. Onillon (corresponding) ---
\author{A.~Onillon\textsuperscript{*}}
%\email{\dagger\,Corresponding author: anthony.onillon@mpi-hd.mpg.de}
\affiliation{IRFU, CEA, Universit\'{e} Paris-Saclay, 91191 Gif-sur-Yvette, France}
\affiliation{Max-Planck-Institut f\"{u}r Kernphysik, 69117 Heidelberg, Germany}

\author{A.~Oralbaev}
\affiliation{APC, Universit{\'e} de Paris, CNRS, Astroparticule et Cosmologie, F-75006, Paris}

\author{C.~Palomares}
\affiliation{Centro de Investigaciones Energ\'{e}ticas, Medioambientales y Tecnol\'{o}gicas, CIEMAT, 28040, Madrid, Spain}

\author{I.M.~Pepe}
\affiliation{Centro Brasileiro de Pesquisas F\'{i}sicas, Rio de Janeiro, RJ, 22290-180, Brazil}

\author{L.~Perisse}
\affiliation{IRFU, CEA, Universit\'{e} Paris-Saclay, 91191 Gif-sur-Yvette, France}

\author{G.~Pronost}
\thanks{Now at: Kamioka Observatory, ICRR, University of Tokyo, Kamioka, Gifu 506-1205, Japan}
\affiliation{Subatech, CNRS, Universit\'{e} de Nantes, IMT-Atlantique, 44307 Nantes, France}

\author{J.~Reichenbacher}
\thanks{Now at: South Dakota School of Mines \& Technology, Rapid City, SD 57701, USA}
\affiliation{Department of Physics and Astronomy, University of Alabama, Tuscaloosa, Alabama 35487, USA}

\author{S.~Sch\"{o}nert}
\affiliation{Physik Department, Technische Universit\"{a}t M\"{u}nchen, 85748 Garching, Germany}

\author{S.~Schoppmann}
\thanks{Now at: Johannes Gutenberg-Universit\"{a}t Mainz, Detektorlabor, Exzellenzcluster PRISMA+, Mainz, Germany}
\affiliation{Max-Planck-Institut f\"{u}r Kernphysik, 69117 Heidelberg, Germany}

\author{L.~Scola}
\affiliation{IRFU, CEA, Universit\'{e} Paris-Saclay, 91191 Gif-sur-Yvette, France}

\author{R.~Sharankova}
\affiliation{Department of Physics, Institute of Science Tokyo, Tokyo, 152-8551, Japan}

\author{V.~Sibille}
\affiliation{IRFU, CEA, Universit\'{e} Paris-Saclay, 91191 Gif-sur-Yvette, France}

\author{V.~Sinev}
\noaffiliation

\author{M.~Skorokhvatov}
\noaffiliation

\author{P.~Soldin}
\affiliation{III. Physikalisches Institut, RWTH Aachen University, 52056 Aachen, Germany}

\author{A.~Stahl}
\affiliation{III. Physikalisches Institut, RWTH Aachen University, 52056 Aachen, Germany}

\author{I.~Stancu}
\affiliation{Department of Physics and Astronomy, University of Alabama, Tuscaloosa, Alabama 35487, USA}

\author{M.R.~Stock}
\affiliation{Physik Department, Technische Universit\"{a}t M\"{u}nchen, 85748 Garching, Germany}

\author{L.F.F.~Stokes}
\affiliation{Kepler Center for Astro and Particle Physics, Universit\"{a}t T\"{u}bingen, 72076 T\"{u}bingen, Germany}

\author{F.~Suekane}
\affiliation{Research Center for Neutrino Science, Tohoku University, Sendai 980-8578, Japan}

\author{S.~Sukhotin}
\noaffiliation

\author{T.~Sumiyoshi}
\affiliation{Department of Physics, Tokyo Metropolitan University, Tokyo, 192-0397, Japan}

\author{C.~Veyssiere}
\affiliation{IRFU, CEA, Universit\'{e} Paris-Saclay, 91191 Gif-sur-Yvette, France}

\author{B.~Viaud}
\affiliation{Subatech, CNRS, Universit\'{e} de Nantes, IMT-Atlantique, 44307 Nantes, France}

\author{M.~Vivier}
\affiliation{IRFU, CEA, Universit\'{e} Paris-Saclay, 91191 Gif-sur-Yvette, France}

\author{S.~Wagner}
\affiliation{APC, Universit{\'e} de Paris, CNRS, Astroparticule et Cosmologie, F-75006, Paris}

\author{C.~Wiebusch}
\affiliation{III. Physikalisches Institut, RWTH Aachen University, 52056 Aachen, Germany}

\author{G.~Yang}
\thanks{Now at: State University of New York at Stony Brook, Stony Brook, NY 11755, USA}
\affiliation{Argonne National Laboratory, Argonne, Illinois 60439, USA}

\author{F.~Yermia}
\affiliation{Subatech, CNRS, Universit\'{e} de Nantes, IMT-Atlantique, 44307 Nantes, France}

% Unindexed corresponding-author line in the title block (no symbol/letter)
\collaboration{\small *\,Corresponding authors:
  \texttt{anthony.onillon@mpi-hd.mpg.de},
  \texttt{thierry.lasserre@mpi-hd.mpg.de}}
\collaboration{Double Chooz Collaboration}

\begin{abstract}
Neutrino emission from nuclear reactors provides real-time insights into reactor power and fuel evolution, with potential applications in monitoring and nuclear safeguards. 
Following reactor shutdown, a low-intensity flux of ``residual neutrinos'' persists due to the decay of long-lived fission isotopes in the partially burnt fuel remaining within the reactor cores and in spent nuclear fuel stored in nearby cooling pools.
The Double Chooz experiment at the Chooz B nuclear power plant in France achieved the first quantitative measurement of this residual flux  based on 17.2 days of reactor-off data.
In the energy range where the residual signal is most pronounced, 
the neutrino detector located 400\,m from the cores recorded $106 \pm 18$ neutrino candidate events (5.9$\sigma$ significance). This measurement is in excellent agreement with the predicted value of $88 \pm 7$ events derived from detailed reactor simulations modeling the decay activities of fission products and incorporating the best-available models of neutrino spectra.
\end{abstract}

\maketitle

\vspace*{-2.59\baselineskip}

Nuclear reactor cores are the largest source of man-made electron antineutrinos ($\bar\nu_e$), produced via the $\beta$-decay of neutron-rich fragments generated primarily during the fission of heavy elements such as uranium and plutonium. Physicists have detected $\bar\nu_{e}$ from reactors for more than six decades~\cite{Cowan:1956rrn}. The emitted $\bar\nu_{e}$'s have unique inherent features that make them of particular interest to International Atomic Energy Agency (IAEA) safeguards, as they are non-alterable and inextricably linked to the nuclear processes occurring in the reactor core~\cite{iaeareport2008}. This concept, pioneered by Borovoi and Mikaelyan in 1978~\cite{Borovoi1978}, was first implemented at the ROVNO power station in 1985~\cite{Afonin:1985rw}. Neutrino emission provides real-time, non-intrusive information about the operational state and fissile content of a reactor core. At leading order, the neutrino flux scales with the total number of fissions, while at the next order, it is influenced by the specific isotopic composition undergoing fission. Neutrino detectors can therefore be used to detect anomalies in neutrino emissions from the cores, indicating possible diversion of nuclear material. Usually, the inverse $\beta$-decay (IBD) capture reaction, $\bar\nu_{e} + p \rightarrow e^{+} + n$, is used for detection. 
Beyond the monitoring of operating nuclear units, several other promising applications of neutrino-based safeguards have been identified through a collaboration between the IAEA and neutrino physics experts~\cite{iaeareport2008}. Among these, the verification of the spent fuel inventory and the estimation of the residual power of the reactor when the core is shut down or following a potential nuclear incident are two topics frequently discussed but not yet quantitatively studied using neutrino data collected during a reactor-off period~\cite{nusafeguards2010,nusafeguards2019}.

After reactor shutdown, a residual neutrino flux continues to be emitted from the decay of long-lived fission products (FPs) present in burnt fuel assemblies still in the reactor core, as well as those previously removed from the cores and stored in nearby spent fuel cooling pools. This residual flux, typically accounting for less than 1\% of the nominal reactor signal, is theoretically well understood but has never been measured. Its detection requires low background conditions, a well-understood detector response, and controlled systematics.
In this paper, we present the first quantitative measurement of this residual neutrino flux and its energy spectrum, obtained with the Double Chooz neutrino experiment~\cite{DoubleChooz:2019qbj}, located in the French Ardennes near the two 4.25\,GW$_{\mathrm{th}}$ cores of the Chooz B nuclear power plant and originally operated to study neutrino oscillations~\cite{Ardellier:2006mn}.

\textit{Post-fission $\bar{\nu}_{e}$.—}The Chooz reactor cores are pressurized water reactors (PWRs), each containing 205 fuel assemblies composed of approximately 600 kg of enriched uranium dioxide (UO$_{2}$), primarily $^{238}$U with a few percent of $^{235}$U. During operation, additional fissile isotopes, $^{239}$Pu and $^{241}$Pu, are produced through neutron capture and subsequent decay processes involving $^{238}$U. The fission of these four isotopes ($^{235}$U, $^{238}$U, $^{239}$Pu, $^{241}$Pu) accounts for more than 99.7\% of the core's thermal power. The reactors typically operate at full power for over a year in what is known as an irradiation cycle, followed by a refueling period during which the reactor is shut down for 6--8 weeks. During this period, about one-third of the spent fuel assemblies are removed and transferred to storage pools located in an adjacent building, approximately 38\,m from the reactor cores.
Each assembly typically undergoes three irradiation cycles and reaches a burnup of approximately 45\,GW$\cdot$days/ton before being removed.
The spent fuel assemblies are then cooled in storage pools for several years before being transported off-site for reprocessing.
Electron antineutrinos ($\bar\nu_{e}$) are mainly produced by the $\beta^{-}$ decay of FPs and, to a lesser extent, by neutron capture reactions forming isotopes such as $^{239}$U and $^{239}$Np. However, only $\bar\nu_{e}$ from FPs are detectable due to the 1.8\,MeV threshold of the IBD reaction. Most fission-induced $\bar\nu_{e}$ are emitted promptly following fission events. During reactor operation, long-lived FPs gradually accumulate in the fuel. After shutdown, these isotopes continue to decay, producing a small residual $\bar{\nu}_e$ flux that decreases over time. Around a reactor, the intensity and spectral shape of this residual flux depend on both the irradiation history and the time elapsed since shutdown, with contributions from fuel assemblies still in the cores and in nearby cooling pools.

\begin{figure}[htbp]
    \centering
    \includegraphics[width=0.4\textwidth]{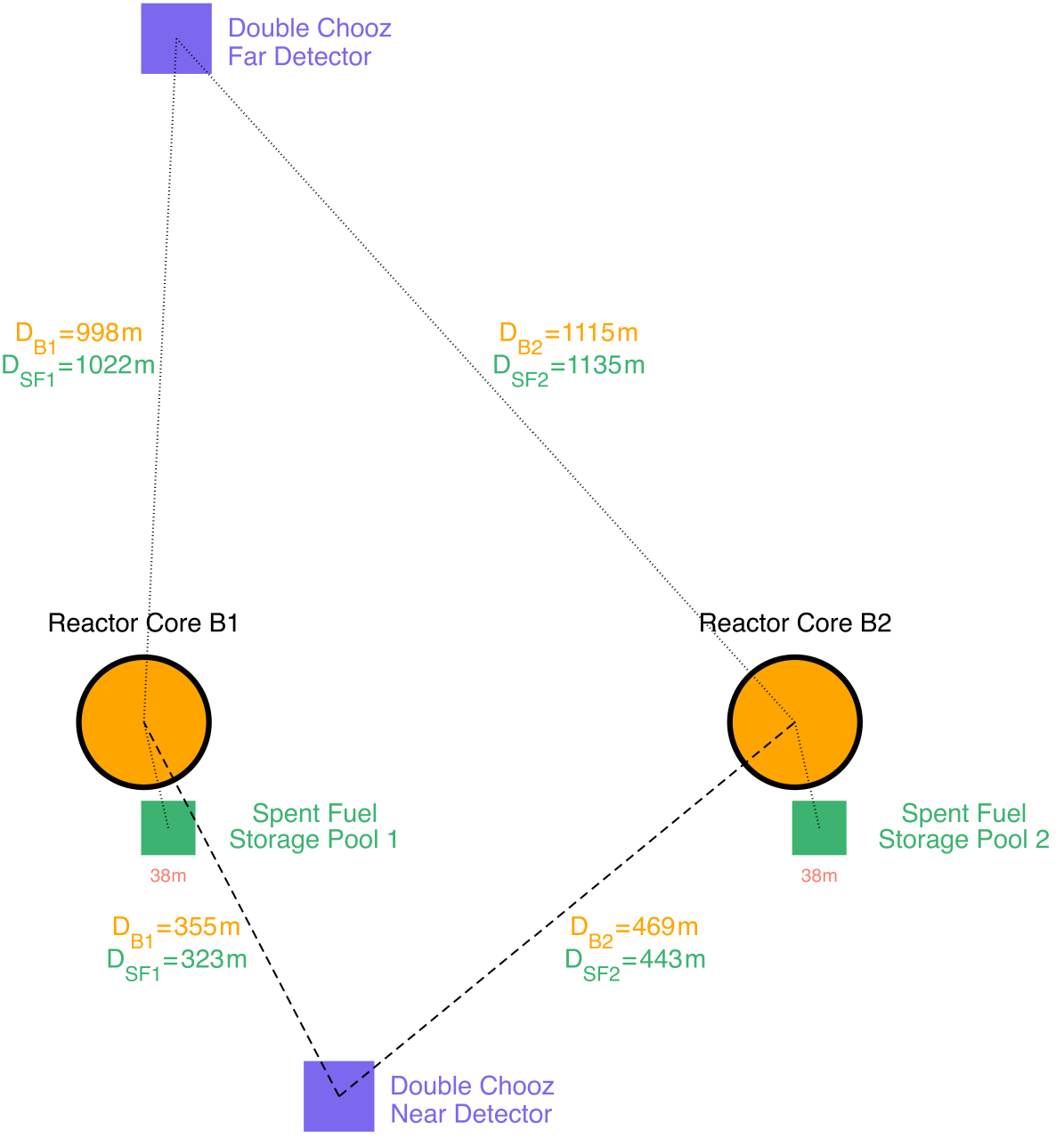}
    \caption{{Schematic of the Double Chooz layout, showing the far and near detectors relative to reactor cores (B1, B2) and their spent fuel pools. }}
    \label{fig:DC_layout}
\end{figure}

\textit{The Double Chooz experiment.—}A two-detector setup was used to measure the $\theta_{13}$ neutrino mixing angle, with detectors placed at average distances of 400\,m and 1.05\,km from the reactor cores to observe neutrino oscillations~\cite{DoubleChooz:2019qbj}. The experiment operated from 2011 to 2017, with the far detector (FD) starting in 2011 and the near detector (ND) joining in 2014 to complete the two-detector setup. Both detectors, shielded underground by 115 (ND) and 300 meters (FD) water equivalent, were identically designed to minimize systematic uncertainties and enable precise near/far comparisons of $\bar\nu_{e}$ rates and energy spectra~\cite{DoubleChooz:2022ukr}. The experimental layout, showing detector positions relative to the reactor cores and spent fuel pools, is illustrated in Figure~\ref{fig:DC_layout}.  
Unlike multi-core reactor experiments such as Daya Bay~\cite{DayaBay:2024hrv} and RENO~\cite{RENO:2020dxd}, where simultaneous core shutdowns never occurred, the two-core configuration of the Chooz B plant enabled in-situ background measurements during such events. This unique feature not only strengthened the original oscillation analysis~\cite{DoubleChooz:2019qbj,DoubleChooz:2020vtr} but also provides a rare opportunity to measure residual $\bar{\nu}_e$ emission.
Each detector consisted of four concentric volumes topped by an outer muon veto system. The innermost volume (\textit{neutrino target}, NT) was filled with 10.3\,m$^{3}$ of Gd-loaded (1\,g/l) liquid scintillator, surrounded by a 22.6\,m$^{3}$ \textit{gamma catcher} (GC) filled with unloaded scintillator to ensure full calorimetry of interactions associated with neutron capture on Gd. These two volumes, along with the 105\,cm-thick buffer tank filled with non-scintillating mineral oil and instrumented with 390 low-background 10-inch photomultiplier tubes (PMTs), formed the \textit{inner detector} (ID). The ID was surrounded by a 50\,cm-thick liquid scintillator \textit{inner muon veto} (IV) equipped with 78 8-inch PMTs. The far detector was shielded from rock radiation by 15\,cm of demagnetized steel, while the near detector was shielded by 1\,m of water. 
An \textit{outer muon veto} (OV), consisting of segmented scintillator modules positioned above the detector, provided additional rejection of cosmic muons. Reactor $\bar\nu_{e}$ were detected via the IBD process, primarily occurring in the NT and GC. At full reactor power, the near and far detectors recorded about 900 and 140 IBD events/day, respectively~\cite{DoubleChooz:2019qbj}.

\begin{figure}[!ht]
    \centering
    \includegraphics[width=0.4\textwidth]{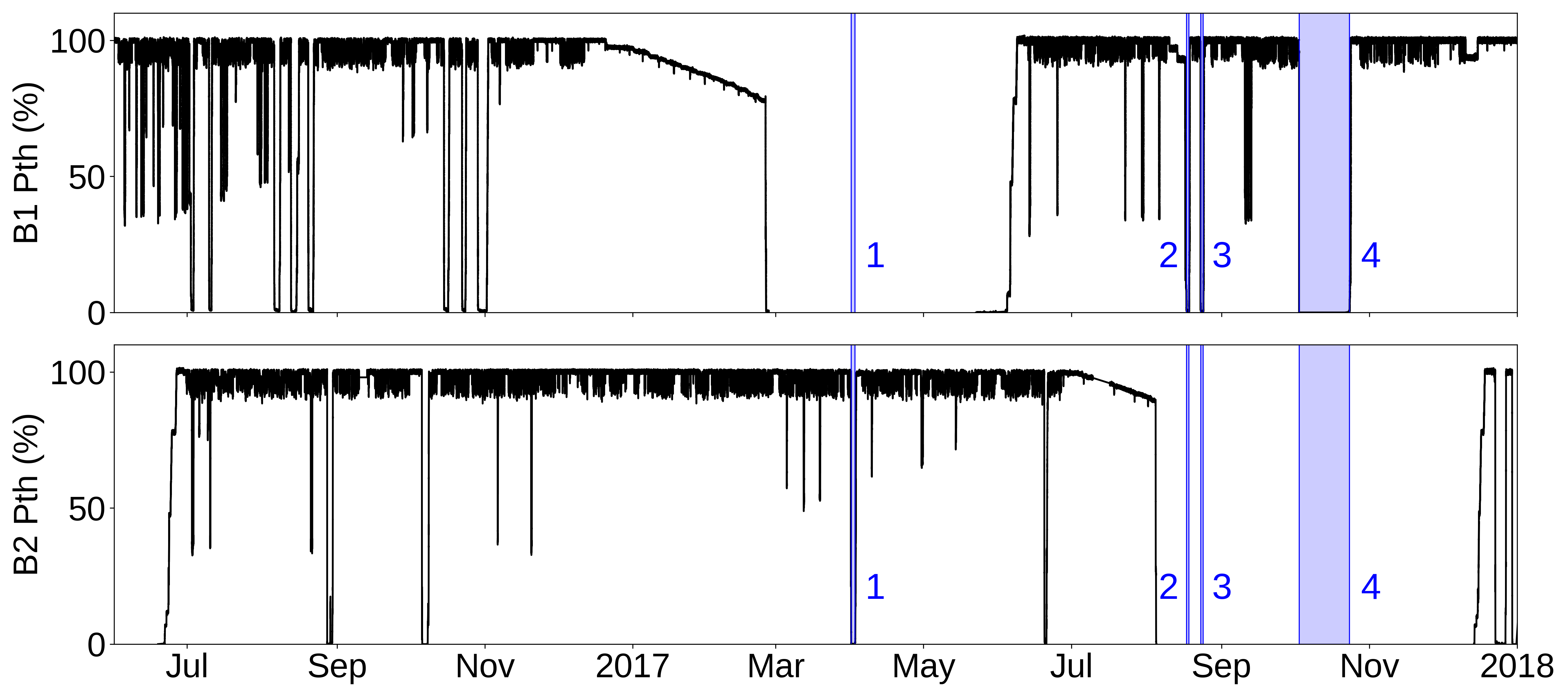}
    \caption{Thermal power history of reactor cores B1 and B2, with the four reactor-off periods in 2017 highlighted in blue.}
    \label{fig:DC_Dataset}
\end{figure}

\textit{Dataset.—}Between 2011 and 2017, the Double Chooz experiment recorded rare reactor-off periods when both reactor cores were simultaneously offline. This scenario was uncommon due to the plant's alternating refueling schedules.
In this work, we focus on a dataset collected in 2017, during four such periods when both reactor cores were simultaneously shutdown for refueling or maintenance. As shown in Figure~\ref{fig:DC_Dataset}, these intervals totaled 24.4 days (1.6, 1.1, 1.0, and 20.8 days).
After accounting for muon veto–induced dead time, the detector livetime amounted to 17.2 days for the near detector and 22.2 days for the far detector.

The analysis was performed for both near and far detectors; however, we focus primarily on the near detector in the following due to its closer proximity to the reactor cores and spent fuel pools, offering higher sensitivity to this low-rate signal.

\textit{Data analysis.—}The IBD detection relies on the characteristic time and spatial correlation between a prompt signal, produced by the positron's ionization ($e^+$) and annihilation, and a delayed signal from neutron (n) capture. The positron's energy deposition directly correlates with the reactor $\bar\nu_{e}$ energy, with $E_{\bar\nu_{e}} \simeq E_{e^{+}} + 0.78 \text{\,MeV}$. IBD event selection follows the same criteria as the oscillation analysis in~\cite{DoubleChooz:2019qbj}, employing the total neutron capture (TnC) method, which, in addition to gadolinium, includes neutron captures on carbon and hydrogen across all ID volumes.
After applying vetoes based on the ID, IV and OV to reduce muon-induced backgrounds, IBD candidates are selected within a 0.5--800\,$\mu$s time window and a spatial separation of 0--1.2\,m between the prompt and delayed signals. The energy windows for the prompt and delayed signals are set to [1, 20]\,MeV and [1.3, 10]\,MeV, respectively. A high-efficiency artificial neural network (ANN) was used to further suppress accidental background contamination by analyzing the time, spatial, and energy correlations of the delayed signals.
Two types of backgrounds mimic the $\bar{\nu}_{e}$ signal: correlated and accidental. 
Due to the relatively shallow detector overburden, correlated backgrounds are dominated by muon-induced processes. 
These arise primarily from fast neutrons generated by muon interactions in the surrounding rock and from $\beta$-n decays of isotopes such as $^{9}$Li, produced by muon spallation on $^{12}$C. 
The rate of $\beta$-n emitters is reduced by tagging their temporal and spatial correlations with the parent muon.
FN rates and energy spectra are estimated using energy depositions tagged by the IV and OV systems up to 20\,MeV. 
Contributions from other sources, such as stopping muons, $\beta$-n decays of $^8$He, and $\beta^{-}$ decay of $^{12}$B, are found to be negligible.
Accidental backgrounds, caused by random coincidences of natural radioactivity and neutron captures, are strongly suppressed by the ANN, with a rejection power exceeding 400. The remaining accidental contamination is measured in-situ by analyzing the rate of single energy depositions within the~ID. 
These selections, applied in the 1--9\,MeV energy range, yield 517 IBD candidates in the near detector, corresponding to a rate of $30.0 \pm 1.4$ events/day. In the far detector, 189 IBD candidates are observed, yielding $8.5 \pm 0.7$ events/day.
The estimated background rates in the same energy window are $26.2 \pm 1.4$ and $7.7 \pm 0.2$ events/day, respectively. 
A summary of the residual $\bar\nu_{e}$ rates, after background subtraction, is provided in Table~\ref{table:IBD_res}. Further details on the detector properties, calibration, data selection, and background estimation are provided in~\cite{DoubleChooz:2019qbj}.

\begin{table}[!h]
\caption{\label{table:IBD_res}%
Background-subtracted and expected residual $\bar\nu_{e}$ rates
for the near and far detectors.}
\begin{ruledtabular}
\begin{tabular}{crr}
Rate (day$^{-1}$) & \multicolumn{1}{c}{ND (400\,m)} & \multicolumn{1}{c}{FD (1.05\,km)} \\ \hline 
\textbf{Data - Background} & & \\
$[$1, 3$]$\,MeV &  6.2\,$\pm$\,1.1   &   1.2\,$\pm$\,0.6      \\
$[$3, 9$]$\,MeV &     $-2.3$\,$\pm$\,1.5 &  $-0.4$\,$\pm$\,0.4   \\\hline
\textbf{Expected Residual $\bar\nu_{e}$} & & \\
$[$1, 3$]$\,MeV &  5.1\,$\pm$\,0.4 & 0.7\,$\pm$\,0.1 \\
$[$3, 9$]$\,MeV & 0.1\,$\pm$\,0.1& 0.1\,$\pm$\,0.1  \\
\end{tabular}
\end{ruledtabular}
\end{table}

\textit{Residual $\bar{\nu}_e$ prediction.—} A detailed summation method model of the residual $\bar{\nu}_e$ rate and energy spectrum emitted from both the reactor cores and the spent fuel pools has been developed for the four reactor-off periods following an approach similar to that in \cite{Kopeikin2001, Kopeikin2012}. This model relies on coupling FP activity predictions with a database of $\bar{\nu}_e$ spectra from $\beta$-decay.

The FP activities are computed using the APOLLO-2.8.4/DARWIN-3 simulation codes~\cite{doi:10.13182/NSE88-3, doi:10.1080/00223131.2000.10875009}, based on the JEFF-3.1.1 nuclear data library~\cite{JEFF-3.1.1}. These widely used and validated tools for neutron transport and residual power calculations simulate the time evolution of the fuel’s isotopic composition and activity during both irradiation phases and cooling periods.

A similar procedure was used to simulate each fuel assembly, whether located in the reactor core or stored in the spent fuel pool during shutdown periods. This involved simulating the individual irradiation history of each assembly up to the beginning of each reactor-off period, including a detailed follow-up of the thermal power it experienced throughout its lifetime as well as any cooling periods between irradiation cycles and after its removal to the spent-fuel pool. An optimized, high-resolution temporal discretization was used to accurately model the buildup and decay of both short- and long-lived fission product isotopes.
Residual $\bar\nu_{e}$ spectra are obtained by coupling the predicted FP activities with the BESTIOLE library~\cite{Bestiole2023}. BESTIOLE models $\bar\nu_{e}$ emission by summing individual $\beta$ branches using an advanced formalism based on Fermi theory, including electromagnetic corrections, finite nuclear size, atomic screening, and shape factors for both allowed and forbidden transitions. 
Particular care is taken in modeling first forbidden non-unique transitions, several of which, such as $^{144}$Pr a major contributor to the residual $\bar{\nu}_e$ flux below 3\,MeV, incorporate detailed nuclear structure calculations.
The IBD spectrum in each detector is obtained by integrating the time-dependent flux over the duration of the reactor-off periods and summing over all sources: 
\begin{align}
\frac{{\rm d}N_{\bar\nu_e}}{{\rm d}E_{\rm vis}} \;&=
  \sum_i \frac{N_{p}}{4\pi L_{i}^{2}} \iint
  \phi_{i}(E,t)\,\sigma_{\rm IBD}(E)\, \notag\\[2pt]
&\quad
  P_{ee}(E,L_{i})\,
  \mathcal{D}\!\bigl(E_{\rm vis}\mid E\bigr)\,
  \mathrm dE\,\mathrm dt, 
\end{align}
where $i$ runs over the two reactor cores and the two spent fuel pools, $N_p$ is the number of target protons, $L_i$ is the baseline to the source $i$, $\phi_i(E, t)$ is the time-dependent energy spectrum of the emitted $\bar{\nu}_e$, $\sigma_{\text{IBD}}(E)$ is the inverse $\beta$-decay cross-section, $P_{ee}(E,L_{i})$ is the survival probability for baseline $L_{i}$, and $\mathcal{D}(E_{\rm vis}\!\mid\!E)$ is the detector response, which folds energy scale, resolution, and selection efficiency to map the true antineutrino energy $E$ onto the reconstructed (visible) energy $E_{\rm vis}$. The IBD cross section formalism of~\cite{PhysRevD.60.053003}, using neutron lifetime value from~\cite{PhysRevD.98.030001} is used. For this calculation, the oscillation parameter $\sin^{2}(2\theta_{13}) = 0.105 \pm 0.014$ from the latest Double Chooz analysis is used~\cite{DoubleChooz:2019qbj}, yielding mean survival probabilities of $0.978 \pm 0.003$ for the near detector and $0.909 \pm 0.012$ for the far detector.

\begin{figure}[!htbp]
    \centering
    \includegraphics[width=0.9\columnwidth]{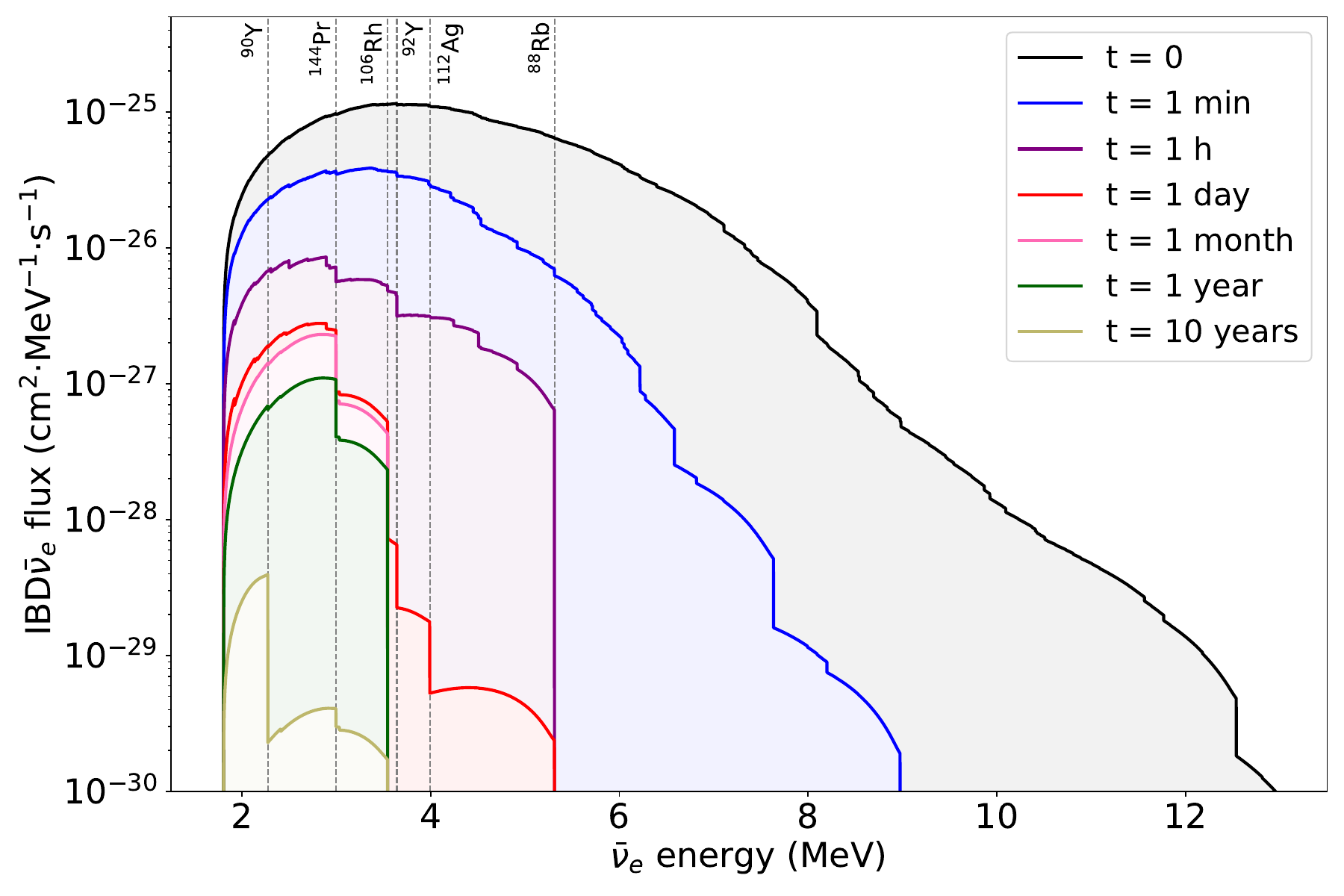}
    \vspace{3mm} % Adjust spacing between the two images
    \includegraphics[width=0.9\columnwidth]{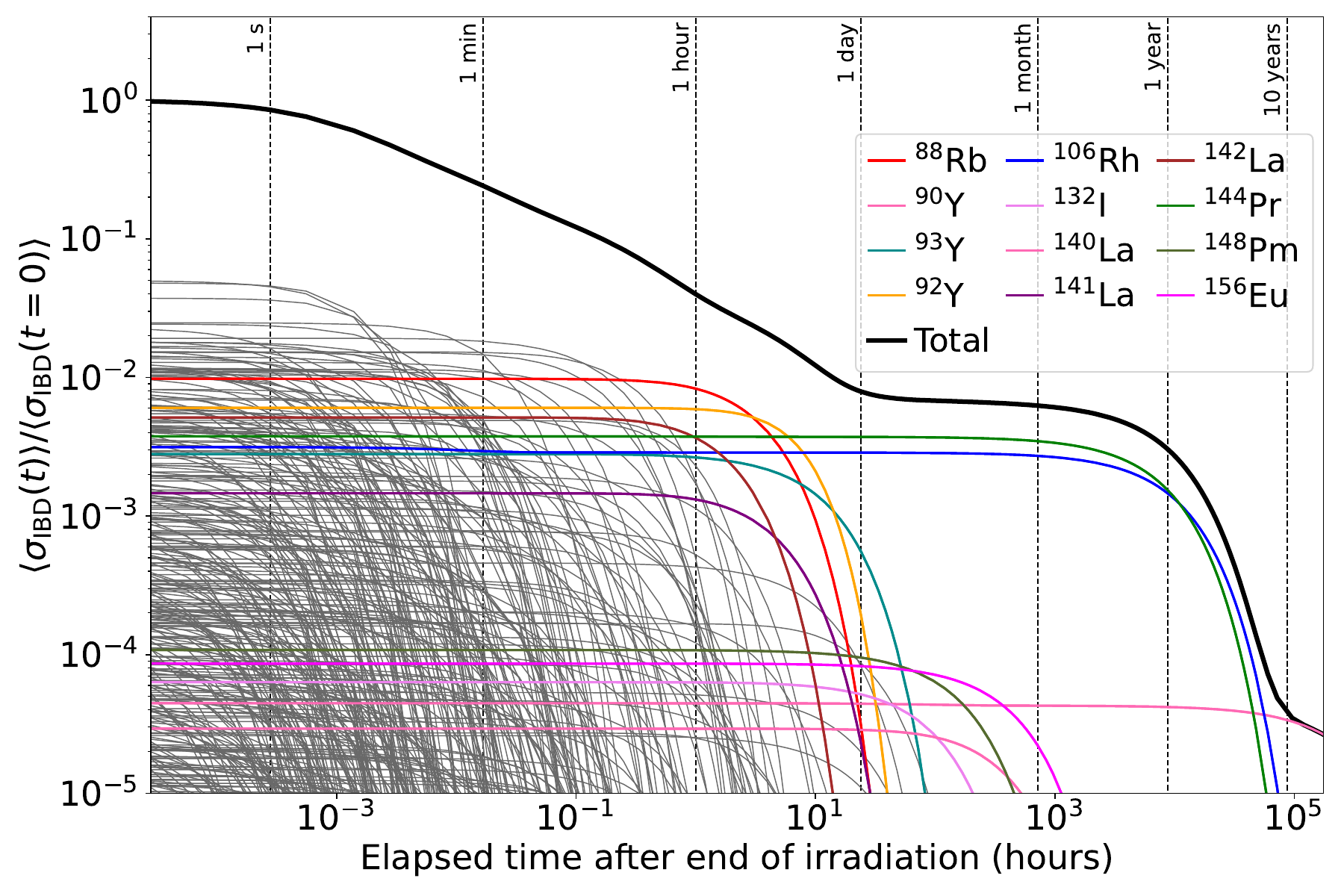}
    \caption{\label{fig:flux_nue_res_assembly} 
    Time evolution of the predicted IBD spectrum (top) and  IBD MCSPF (bottom) for a UO$_2$ (4\,wt\% enrichment) spent fuel assembly irradiated to 45\,GWd/t. Isotope contributions to the MCSPF are gray; dominant ones after one hour are colored.}
\end{figure}

To first order, the rate and energy spectrum of residual $\bar{\nu}_e$ emitted by a burnt fuel assembly are primarily determined by the burnup achieved before shutdown, which sets the inventory of accumulated fission products.
As shown in Figure~\ref{fig:flux_nue_res_assembly} for a UO$_{2}$ assembly irradiated to 45 GWd/t, the predicted IBD $\bar\nu_{e}$ spectrum exhibits a rapid decline in both intensity and mean energy following reactor shutdown. 
As short-lived isotopes decay, the IBD mean cross-section per fission (MCSPF) decreases by one, two, and three orders of magnitude after $\sim$10\,min, $\sim$15\,h, and $\sim$2.5\,y, respectively. Over the same period, the mean $\bar{\nu}_e$ energy drops from $\sim$4.2\,MeV to $\sim$3.2\,MeV, and eventually to $\sim$2.7\,MeV.
Immediately after shutdown, the flux results from the superposition of hundreds of $\beta$ branches from many fission products. 
Within a few hours, however, it becomes dominated by a small number of longer-lived isotopes. 
After several months, the primary contributors are the short-lived isotopes $^{90}$Y ($T_{1/2} = 3.19$\,h, $Q_{\beta} = 2.28$\,MeV), $^{106}$Rh (30.1\,s, 3.54\,MeV), and $^{144}$Pr (17.3\,min, 3.00\,MeV), whose long-lived precursors have accumulated in the fuel: $^{144}$Ce–$^{144}$Pr ($T_{1/2} = 285$\,d), $^{106}$Ru–$^{106}$Rh (372\,d), and $^{90}$Sr–$^{90}$Y (28.9\,y). Beyond ten years, $^{90}$Y alone accounts for more than 90\% of the residual flux.

\textit{Reactor and pool contributions.—}Burnt fuel assemblies remaining in the reactor core during refueling or maintenance are a significant source of residual $\bar\nu_{e}$ in the early stages of a reactor-off period, with the flux strongly influenced by short-lived isotopes within the first tens of hours after shutdown.
In contrast, spent fuel stored in cooling pools contributes a long-lived component to the residual flux. Unlike burnt fuel, this contribution arises from the cumulative emission of assemblies removed over multiple fuel cycles. Some have been cooling for years, while others were removed more recently, resulting in a broad distribution of cooling times and activity levels that shapes both the intensity and spectrum of the emitted $\bar{\nu}_e$.
 
\begin{figure}[!ht]
    \centering
    \includegraphics[width=0.9\columnwidth]{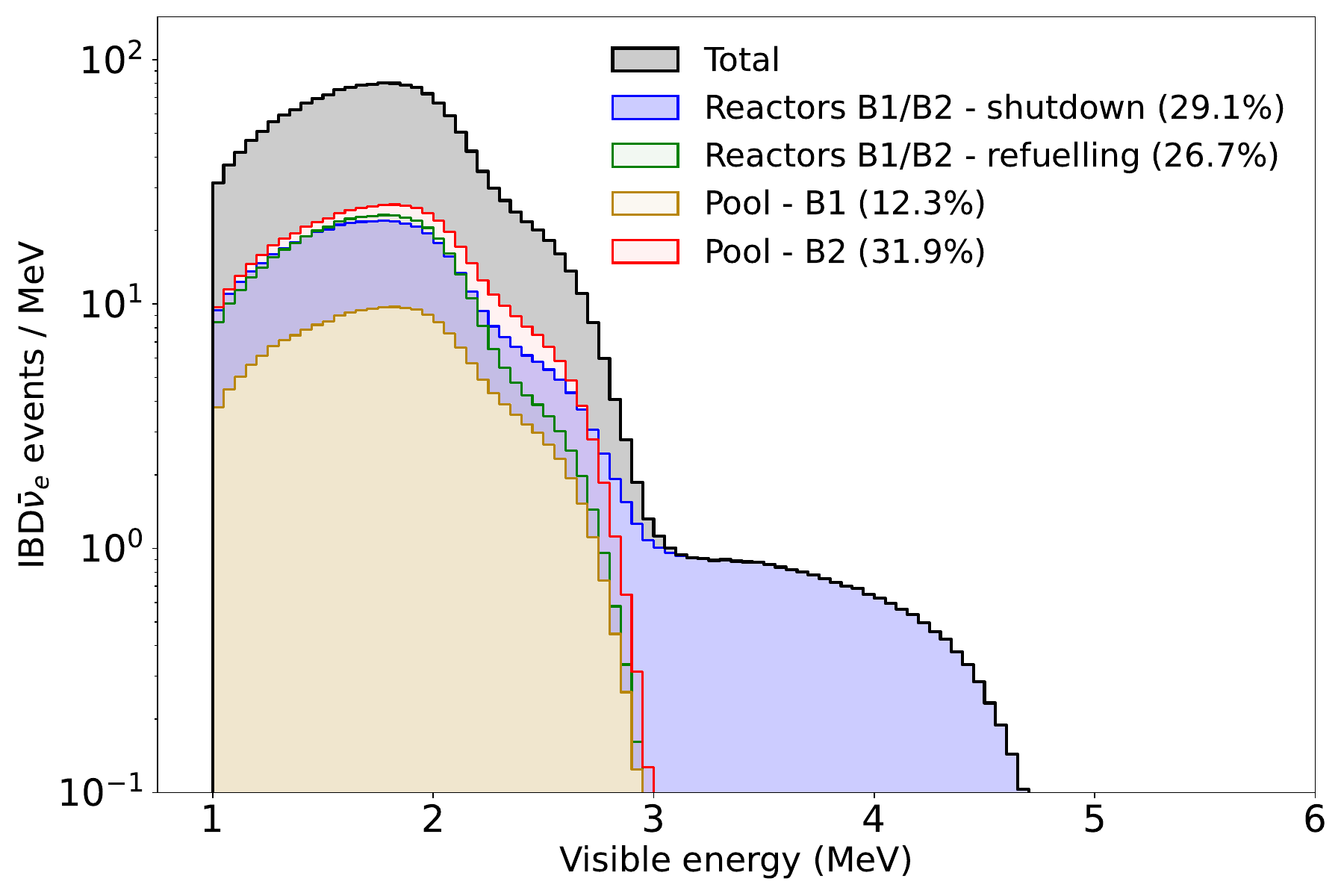}  
    \caption{\label{fig:ND_IBD_spectrum_Pred} 
    Predicted IBD spectrum in the near detector for all off-off periods stacked and combined, showing contributions from two reactor cores, two spent fuel pools, and the total spectrum. 
    }
\end{figure}

Figure~\ref{fig:ND_IBD_spectrum_Pred} shows the predicted residual $\bar\nu_{e}$ IBD spectrum for the Double Chooz near detector, including contributions from both reactor cores and spent fuel pools. 
The predicted IBD spectrum extends up to approximately 4.5\,MeV with 98.7\% of the residual flux expected below 3\,MeV. An energy-integrated contribution of 56\% from the reactors and of 44\% from the storage pools is obtained. 
While residual $\bar{\nu}_e$ emission is initially dominated by burnt fuel still in the reactors due to the presence of short-lived isotopes, the long-term emission from spent fuel pools, integrating multiple assemblies with varying cooling times, becomes comparable over extended reactor-off periods.
The B1 pool (yellow area) contributes less than B2 (red area) because the longest reactor-off period coincided with the refueling of reactor B2. During this time, a new batch of spent fuel was added to the B2 pool (red line), increasing its relative contribution to the total residual $\bar\nu_{e}$ flux.
The total normalization uncertainty on the predicted residual $\bar{\nu}_e$ signal in the near detector is 7.4\%, dominated by a 6.0\% uncertainty in antineutrino spectrum modeling. This uncertainty arises primarily from nuclear-structure uncertainties affecting the dominant first-forbidden transitions of $^{144}$Pr. Conservatively, the 1$\sigma$ uncertainty on the $^{144}$Pr spectral shape was defined as the difference between the shape factor obtained from the detailed nuclear-structure calculations and that obtained using the simplified $\xi$-approximation, which assumes an allowed transition.
Other significant contributions come from geometric baseline uncertainties (2.9\%), the simulated fission product inventory (2.1\%), and the estimated amount of spent fuel assemblies in the pools (2.0\%). The latter includes a conservative treatment of older assemblies whose status is uncertain. Detector-related systematics are subdominant, contributing less than 0.8\% in total.

\begin{figure}[!htbp]
\includegraphics[width=0.9\columnwidth]{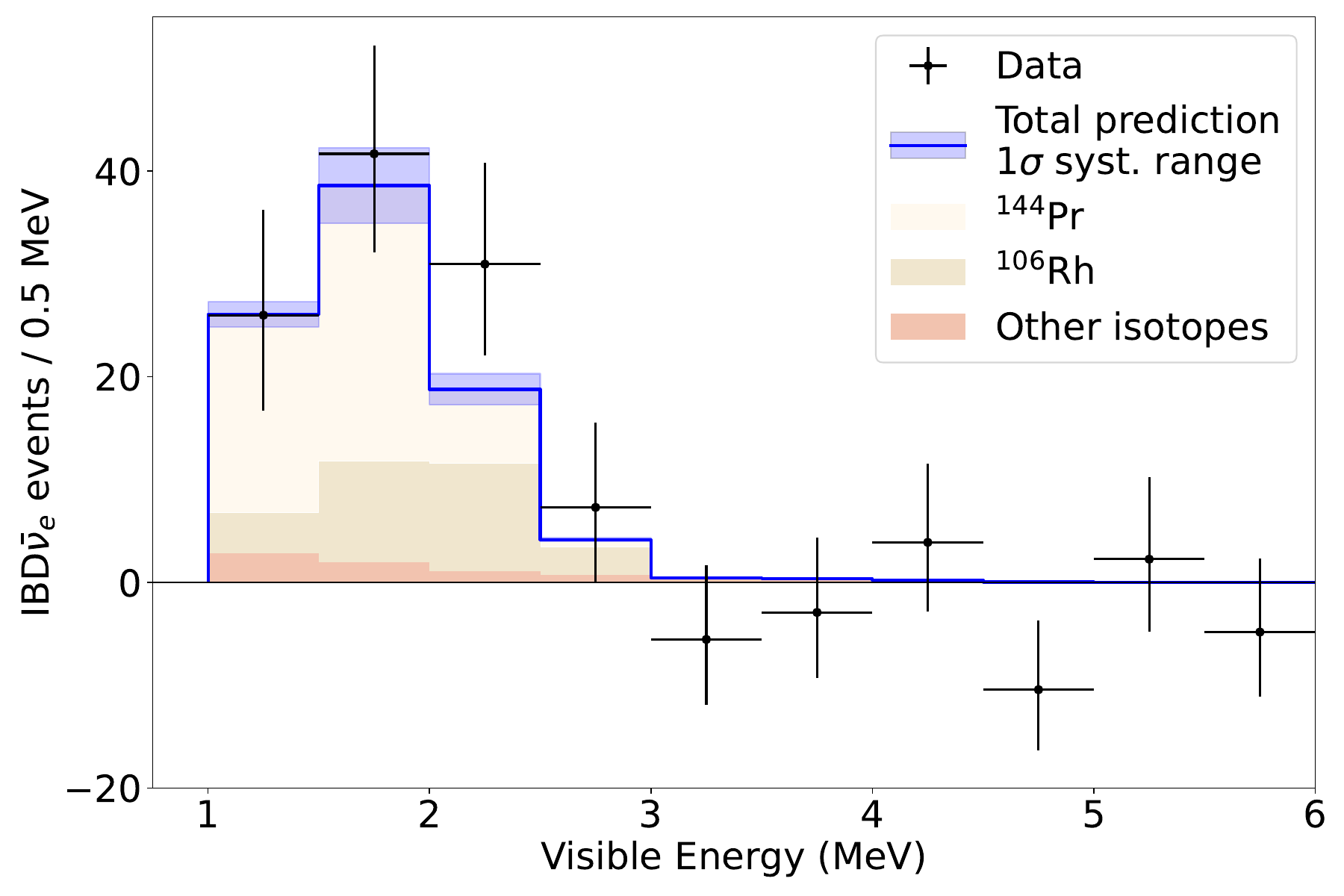}  
\caption{Measured residual $\bar\nu_{e}$ spectrum in the Double Chooz near detector (background subtracted), compared with the prediction, highlighting contributions from the dominant isotope contributors. 
In the 1--3\,MeV range, $106 \pm 18$ events are observed, compared to a prediction of $88 \pm 7$ events.} 
\label{fig:ND_IBD_spectrum} 
\end{figure}

\textit{Results.—}Measured and predicted residual $\bar\nu_{e}$ rates are summarized in Table~\ref{table:IBD_res} for different energy ranges. In the near detector, a significant excess of IBD events is observed in the 1–3 MeV window, where the residual signal is expected to peak, as clearly shown in Figure~\ref{fig:ND_IBD_spectrum}. In this range, $106 \pm 18$ residual candidates are observed, in good agreement with the prediction of $88 \pm 7$ ($\Delta = 18 \pm 19$), both in rate and in spectral shape. 
The measurement uncertainty is dominated by statistical fluctuations in the observed IBD candidates, with a total of 244 events recorded before background subtraction.
The flux in this region is expected to be dominated by $^{144}$Pr ($\sim$54\%) and $^{106}$Rh ($\sim$38\%), with minor isotopes accounting for the remainder.
In the far detector, a smaller excess of $27 \pm 13$ events is observed in the same energy range, compared to the predicted $14 \pm 1$. Although this measurement is also in good agreement with the prediction, the larger baseline suppresses both the event rate and the statistical significance relative to the near detector.
Above 3\,MeV, no significant excess is observed in either detector. In this region, the background-subtracted yields are consistent with zero (see Table \ref{table:IBD_res}), validating the background model and confirming that the small high‑energy residual tail, predicted at a rate below 0.1 events/day, lies well below the current sensitivity.

\textit{Conclusion.—}A first analysis of 7.5 days of reactor-off data collected in 2011 and 2012 using only the far detector validated the predicted background model for the neutrino oscillation analysis~\cite{DoubleChooz:2012vqd}. However, the short reactor-off period, along with the absence of near detector data, limited the sample size to only $\sim$20 IBD candidates, which was insufficient to extract a residual reactor $\bar{\nu}_{e}$ spectrum.
The present work overcomes these limitations by using 17.2\,days of reactor-off data collected in 2017 with the near detector. 
Combined with a detailed summation-model prediction based on nuclear fuel simulations, this dataset enables the first quantitative measurement of residual reactor $\bar{\nu}_{e}$ emission and the first direct validation of residual-flux calculations. 
In the 1--3\,MeV range, where the residual $\bar{\nu}_{e}$ signal is the strongest, the observed $106 \pm 18$ events represent a 5.9$\sigma$ excess over background, in excellent agreement with the predicted $88 \pm 7$ events.
This measurement establishes a proof of principle for $\bar{\nu}_{e}$-based monitoring in the context of nuclear safeguards, demonstrating that neutrino detectors can provide direct and non-intrusive insight into reactor activity, even during shutdown. It paves the way for new applications in the verification of spent fuel inventories and the development of long-term reactor monitoring strategies.
\textit{Acknowledgments.—}
We thank the company {\it Electricit\'e de France} (EDF); 
the European fund FEDER; 
the R\'egion de Champagne-Ardenne; 
the D\'epartement des Ardennes;
and the Communaut\'e de Communes Ardenne Rives de Meuse.
We acknowledge the support of 
%France
the CEA, 
CNRS/IN2P3, 
the computer centre CC-IN2P3,
and 
LabEx UnivEarthS in France;
%Germany
the Max Planck Gesellschaft, 
the Deutsche Forschungsgemeinschaft DFG, 
the Transregional Collaborative Research Center TR27, 
the excellence cluster ``Origin and Structure of the Universe'',
and 
the Maier-Leibnitz-Laboratorium Garching in Germany;
%Japan
the Ministry of Education, Culture, Sports, Science and Technology of Japan (MEXT),
and
the Japan Society for the Promotion of Science (JSPS) in Japan;
%Spain
the Ministerio de Econom\'ia, Industria y Competitividad 
%(SEIDI-MINECO) under grants FPA2016-77347-C2-1-P and MdM-2015-0509 
in Spain;
%USA
the Department of Energy and the National Science Foundation in the United States;
%Russian
the Russian Academy of Sciences, 
the Kurchatov Institute, 
and 
the Russian Foundation for Basic Research (RFBR) in Russia;
%Brasil
the Brazilian Ministry of Science, Technology and Innovation (MCTI), 
the Financiadora de Estudos e Projetos (FINEP), 
the Conselho Nacional de Desenvolvimento Cient\'ifico e Tecnol\'ogico (CNPq), 
the S\~ao Paulo Research Foundation (FAPESP)
and 
the Brazilian Network for High Energy Physics (RENAFAE) in Brazil.

\bibliographystyle{apsrev4-2}
\bibliography{3-bibliography} 
 
\end{document}